\documentclass[12pt]{article}

\def\beq{\begin{equation}}
\def\eeq{\end{equation}}
\def\barr{\begin{eqnarray}}
\def\beqa{\begin{eqnarray}}
\def\earr{\end{eqnarray}}
\def\eeqa{\end{eqnarray}}
\def\winf{W_{1+\infty}\ }
\def\u1{\widehat{U(1)}}
\def\v{V\,}
\def\w{W\,}
\def\vb{{\overline V}\,}
\def\wb{{\overline W}\,}

\newcommand{\nl}{\nonumber \\}

\begin{document}

\begin{titlepage}

\begin{center}
\hfill  \quad  \\
\vskip 0.5 cm
{\Large \bf Effective short distance interaction in Calogero-Sutherland quantum fluids}

\vspace{0.5cm}

Federico~L.~ BOTTESI$^a$ ,\ \ Guillermo~R.~ZEMBA$^{b,}$\footnote{
Fellow of Consejo Nacional de Investigaciones Cient\'{\i}ficas y T\'ecnicas,Argentina.}\\

{\em $^a$Facultad de Ingenier\'ia, Universidad de Buenos Aires,}\\
{\em  Av. Paseo Col\'on 850,(C1063ACL) Buenos Aires, Argentina}\\
\medskip
{\em $^b$Departamento de F\'{\i}sica Teórica,GIyA,Laboratorio Tandar,}\\
{\em  Comisi\'on Nacional de Energ\'{\i}a At\'omica,} \\
{\em Av.Libertador 8250,(C1429BNP) Buenos Aires, Argentina}
{\em and}\\
{\em Facultad de Ingenier\'ia y Ciencias Agrarias,  Pontificia Universidad Cat\'olica Argentina,}\\
{\em  Av. Alicia Moreau de Justo 1500,(C1107AAZ) Buenos Aires, Argentina}\\

\medskip

\end{center}
\vspace{.3cm}
\begin{abstract}
\noindent
We consider the effective conformal field theory with symmetry $\winf \times {\overline \winf}$ that describes
the thermodynamic limit of the Calogero-Sutherland model. 
In the repulsive regime of the free fermion formulation, we identify an attractive interaction 
between opposite moving particle-hole pairs that 
dominates the short distance behavior and that is proposed as responsible for the destabilization of the ground
state, leading to a new one of bosonic nature. The process is described by a Bogoliubov transformation of the free 
fermion bilinear operators into bosonic ones, preserving the form of the $\winf$ algebra but decoupling the 
opposite chirality terms in the hamiltonian, as expected in the low energy limit.
In coordinate space this interaction has a short range component that arises due to the quantum regularization of the theory. 
The described dynamical process 
may be considered as a mechanism of the emergence of the known charge and quantum statistics fractionalization
of the low lying excitations of the theory, as predicted in both first and second
quantization studies. 
\end{abstract}
\vskip 0.5 cm
\end{titlepage}
\pagenumbering{arabic}
The Calogero-Sutherland (CS) model \cite{cal,sut} of non-relativistic fermions has been studied in both its first 
and second quantized forms, leading
to interesting physical phenomena, like charge and statistics fractionalization (anyonic character
of its low energy excitations) \cite{hald1,cs-reviews} as well as quantum fluidity of the Benjamin-Ono (BO) class 
\cite{boreview} in the thermodynamic limit \cite{aw,abw,bz}. Due to the first property, it has been proposed as candidate model
for describing the Laughlin fluids in the quantum Hall effect (QHE) \cite{laugh}, as well as other similar
systems. This identification, however, faces two difficulties: {\it i)} the need for a chiral projection 
(due to the magnetic field in the QHE) that is absent in the original, parity preserving theory 
and, {\it ii)} the compressible character of the corresponding quantum fluid, which does not match
the incompressibility of the Laughlin theory. Actually, the compressibility $\kappa$ 
of the CS quantum fluid may be calculated from the expression $1/\kappa \propto d \mu /d N$,
where $\mu$ is the chemical potential found in \cite{clz,flsz} and $N$ the particle number,
yielding $\kappa \propto 1/{\xi}^2 $, where $\xi$ is a reparametrization of the coupling 
constant of the CS model $g$, as defined momentarily. 
Therefore, the CS quantum fluid describes a gas, rather than a liquid, that sustains density waves of the 
Benjamin-One class (solitons and anti solitons of width governed by $\xi$). It may describe a 
1D section of the 2D bulk (magneto-rotons) or a higher temperature phase of
the quantum  incompressible fluid, sharing with it the charge and statistics fractionalization, 
but this is left as an open issue. The origin of the anyonic character is still, to the best of our knowledge,
an open question. This property is expected to be the consequence of a dynamical process due to a strong attractive
interaction between pairs of fermions that destabilize the free fermion vacuum inducing a new type
of ground state, in which low lying excitations are anyons. 
The origin of the special properties of the CS model may be traced down to the peculiar 
properties of the $ 1/r^2$ interactions \cite{lanlif} in 3D, from which we list just two:

{\it i)} It resembles a 'centrifugal barrier' $U_L(r)=L^2/(2mr^2)={\hbar}^2\ell (\ell +1 )/(2mr^2)$ in 3D 
systems subject to a central potential, such that orbital angular momentum $L$ is conserved and the resulting planar motion
may be rewritten in terms of the radial evolution only ($\ell$ is the quantum number of orbital angular momentum).

{\it ii)} For dimensional reasons, it yields the only scale-free potential in quantum mechanics, in units such that $\hbar = 1$

Recent related work in vortex matter shows the appereance of edge waves on a
 boundary layer of vorticity whose dynamics is governed by the BO equation
\cite{bw}. Similar results have been reported for the quantum hydrodynamics 
 of phonons in 1D \cite{prlam}. Some recent research on the topics of this letter has been reported in \cite{km,moll,issch}.

In the following we will be focusing on the Sutherland model, a 1D theory with
pairwise interaction potential proportional to the inverse square distance in 3D space.
Consider a system of $N$ non-relativistic 
$(1+1)$-dimensional spinless fermions on a circumference of length $L$, with 
Hamiltonian \cite{sut} (in units where $\hbar =1$ and $2m=1$, with $m$ being 
the mass of the particles)
\beq
h_{CS}=\sum_{j=1}^N\ \left( \frac{1}{i} \frac{\partial}{\partial x_j}\
\right)^2\ +\ g\ \frac{\pi^2}{L^2}\ \sum_{i<j}\ \frac{1}{\sin^2
(\pi(x_i-x_j)/L) }\ ,
\label{ham}
\eeq
where $x_i$ ($i=1,\dots,N$) is the coordinate of the $i$-th particle
along the circumference, and $g$, 
the dimensionless coupling constant. Ground state stability imposes $g \geq -1/2$,
with both attractive ($-1/2 \leq g < 0$) and repulsive ($0 < g$) regimes. A standard 
reparametrization of the coupling constant is given by $g=2 \xi ( \xi -1)$, 
so that $\xi \geq 0$, where $0 \leq \xi < 1$ is the attractive region and $1  < \xi $ 
the repulsive one. Note that the parameter can be rewritten as $\xi = \ell +1$, 
where $\ell$ can be interpreted as the quantum number of orbital angular momentum.
In the following, we shall consider the repulsive regime because it is 
the one relevant for charge and statistics fractionalization, in the sense
of the QHE. 

Consider now the thermodynamic limit of the CS model (\ref{ham}). Our description is based on the
effective field theory \cite{polch} obtained in \cite{clz,flsz,cfslz}\ by
reformulating the system dynamics in terms of fields
describing the low energy fluctuations of the 1D Fermi surface (two isolated points, actually).
The suitable fermionic fields are obtained from non-relativistic fermionic free fields, 
and the EFT is obtained by taking the thermodynamic limit $N \to \infty$ properly on the fields and hamiltonian.
Two sets of relativistic Weyl fermion fields are obtained around each of the Fermi points.
In the last step, the hamiltonian is rewritten in terms of new fields that obey 
the $\winf$ algebra \cite{shen,kac1}. It is a key step in our approach, as it allows to diagonalize the 
hamiltonian and find the Hilbert space of the EFT by using to good advantage 
the algebraic properties of the $\winf$ algebra. Bosonization of the interacting fermion fields \cite{boson}
may be achieved by finding bosonic realizations of the algebra. The entire procedure has been termed 
{\it algebraic bosonization}.
The structure of the Hilbert space is
obtained in the framework of extended conformal field theory (CFT) \cite{bpz}.
The relevant theories are $(c,{\overline c})=(1,1)$ CFTs of a compactified free boson field with extended 
symmetry $\winf \times {\overline \winf}$ and chiral and
antichiral sectors that are isomorphic ($c$ and ${\overline c}$ are the central charges of the chiral and
antichiral sectors). The Hilbert space and partition
function for these theories are known and their dynamics is given by the specific CS
hamiltonian \cite{bz}.
Alternative descriptions of the effective field theory may be found in \cite{kaya,khve,amos,poly}.

The general form of the $\winf$ algebra satisfied by the operators $V^j_m$, with $j=0,1,2,\dots$ and $m$ integer
is:
\beq
\left[\ V^i_\ell, V^j_m\ \right] = (j\ell-im) V^{i+j-1}_{\ell+m}
+q(i,j,\ell,m)V^{i+j-3}_{\ell+m}
+\cdots +\delta^{ij}\delta_{\ell+m,0}\ c\ d(i,\ell) \ ,
\label{walg}
\eeq
where the structure constants $q(i,j,\ell,m)$ and $d(i,\ell)$ 
are polynomial in their arguments, $c$ is the central charge, 
and the dots denote a finite number of terms involving the operators 
$V^{i+j-2k}_{\ell+m}\ $.
The $c=1$ $\winf$ algebra can be realized by either fermionic or bosonic 
operators. The first realization is useful for identifying the correct hamiltonian
operator content, and is given by:
\barr
\v^0_n &=& \sum_{r=-\infty}^{\infty}: a^\dagger_{r-n}\, a_r :
{}~~~,\nl
\v^1_n &=& \sum_{r=-\infty}^{\infty} \left(\,r-{n+1\over 2}\,\right)
:  a^\dagger_{r-n}\, a_r :~~~,\label{fockw}\\
\v^2_n &=& \sum_{r=-\infty}^{\infty} \left(\,r^2 -(n+1)\ r +
{{(n+1)(n+2)}\over 6}\,
\right) :  a^\dagger_{r-n}\, a_r :~~~,\nl
\nonumber
\earr
with $\{\ a_k , a^{\dag}_l\ \}\ =\ \delta_{k,l}\ $ and all other 
anticommutators vanishing. These are explicit expressions for the right ($R$) 
Fermi point with analogous ones for the left ($L$) one.
In obtaining these results, the standard procedure of considering a shell of 
$O(\sqrt{N})$ momentum modes around each Fermi point, and then taking the 
$N \to \infty$ limit has been applied \cite{flsz} . 
The fermionic ground state $|\ \Omega\ \rangle_0$ is a highest-weight state 
with respect to the $\winf $ operators, {\it i.e.},
$V^i_\ell |\ \Omega\ \rangle_0 = 0$, $\ell > 0, i \geq 0$,
which expresses the incompressibility of the Fermi sea.
The operators defined in (\ref{fockw}) satisfy the $\winf$ algebra.

The $\winf$ algebra may be realized in terms of a chiral bosonic field 
by a generalized Sugawara construction \cite{kac1} .
In fact,defining the right and left moving modes,
$\alpha_\ell$ and ${\overline \alpha}_\ell$, of a free compactified 
boson ($[\alpha_n,\alpha_m ]=\xi n \delta_{n+m,0}$ and similarly for 
the ${\overline \alpha}_\ell$ 
operators), the commutation relations (\ref{walg1}) 
are satisfied by defining $\w^i_\ell$ (we only write the expressions
for $i=0,1,2$) as:
\barr
\w^0_\ell &=& \alpha_\ell ~~~,
\nonumber\\
\w^1_\ell &=& {\frac{1}{2}} \sum_{r= -\infty}^{\infty}
:\, \alpha_{r}\,\alpha_{\ell-r}\,
:~~~,\label{mod2}\\
\w^2_\ell &=& {\frac{1}{3}} \sum_{r, s = -\infty}^{\infty}
:\, \alpha_{r}\,\alpha_s\, \alpha_{\ell-r-s}\,:~~~,
\nonumber
\earr
and analogously for the operators $\wb^i_\ell$ in terms of ${\overline \alpha}_\ell$.
The naive generalization of (\ref{mod2}) to higher 
values of conformal spin is incorrect (for example, see \cite{flsz}).

For the case of the CS EFT, $c=1$, and all the relevant commutation relations are:
\barr
\left[\ \w^0_\ell,\w^0_m\ \right] & = &  c\ \xi \ell\ \delta_{\ell+m,0} ~~~,\nl
\left[\ \w^1_\ell, \w^0_m\ \right] & = & -m\ \w^0_{\ell+m} ~~~,\nl
\left[\ \w^1_\ell, \w^1_m\ \right] & = & (\ell-m)\w^1_{\ell+m} + 
\frac{c}{12}\ell(\ell^2-1) \delta_{\ell+m,0}~~~,\nl
\left[\ \w^2_\ell, \w^0_m\ \right] &=& -2m\ \w^1_{\ell+m}~~~,
\label{walg1}\\
\left[\ \w^2_\ell, \w^1_m\ \right] &=& (\ell-2m)\ \w^2_{\ell+m} -
   \frac{1}{6}\left(m^3-m\right) \w^0_{\ell+m}~~~,\nl
\left[\ \w^2_n, \w^2_m\ \right] &=& (2n-2m)\ \w^3_{n+m}
     +{n-m\over 15}\left( 2n^2 +2m^2 -nm-8 \right) \w^1_{n+m}\nonumber\\
     &&\quad +\ c\ {n(n^2-1)(n^2-4)\over 180}\ \delta_{n+m,0}~~~.\nonumber
\earr
The first and third equations in (\ref{walg1}) show
that the generators $\w^0_\ell$ satisfy the 
Abelian Kac-Moody algebra $\u1$, and the generators $\w^1_\ell$ 
the Virasoro algebra, respectively.
The operators $\wb^i_\ell$ obbey the same algebra (\ref{walg1}) 
with central charge ${\overline c}=1$ 
and commute with the all the operators $\w^i_\ell$. For this reason, the complete EFT 
of the CS model is a $(c,{\overline c})=(1,1)$ CFT, but since both chiral and
antichiral sectors are isomorphic, we will often consider one of them for
simplicity.

The effective hamiltonian of the EFT of the CS model may be given as power series in $1/N$ in the 
fermionic $\winf$ basis 
according to \cite{cfslz,flsz}:
\beq
{\cal H}_{CS} = \sum_{k=0}^\infty
\left(\frac{2\pi}{N}\right)^k\,{\cal H}_{(k)}~~~, 
\label{series}
\eeq
where
\beq
{\cal H}_{(0)} =  \frac{1}{4} (1+g) \left(\v^0_0+
\vb^0_0\right)~~~,
\label{hcv0}
\eeq
\beq
{\cal H}_{(1)} = \left(1+\frac{g}{2}\right) \left(\v^1_0+
\vb^1_0\right)+\frac{g}{2} \sum_{\ell=-\infty}^{\infty}
\v^0_{\ell}\, \vb^0_{\ell}~~~,
\label{hcv1}
\eeq
\barr
{\cal H}_{(2)} &=&
\left(1+\frac{g}{4}\right) \left(\v^2_0+\vb^2_0\right)
-\frac{1}{12} (1+g) \left(\v^0_0+\vb^0_0\right) \nl
&&-\ \frac{g}{4}\sum_{\ell=-\infty}^{\infty} |\ell |
\left( \v^0_{\ell}\,\v^0_{-\ell}+
\vb^0_{-\ell} \,\vb^0_{\ell}+
2\,\v^0_{\ell} \,\vb^0_{\ell}\right)\nl
&&+\ \frac{g}{2}
\sum_{\ell=-\infty}^{\infty} \left( \v^1_{\ell}\,\vb^0_{\ell}+
\v^0_{\ell} \,\vb^1_{\ell} \right) ~~~.
\label{hcv2}
\earr
The last term includes a four-fermion interaction that has diverse
pieces. 
Consider the middle term in (\ref{hcv2}). Notice that it provides 
an {\it attractive} interaction. The coefficient $|\ell |$ is the 
matrix element of the four-body interaction between 
pairs of fermion bilinears in momentum space. 
It grows linearly with
the exchanged momentum $\ell$, which is also a measure of 
orbital angular momentum in the spatial domain. 
This exchange may occur between particle-hole pairs
on the same Fermi point as well as in backward scattering
processes. Given that all other matrix elements in (\ref{hcv2})
are constant, we infer that this attractive interaction dominates
the short distance, high energy dynamics of the system. 
The free fermion ground state may become unstable
under the effect of this interaction.
From the conceptual point of view, we regard the EFT as defined by the Hilbert
space and the hamiltonian (\ref{series}) as the starting point of our analysis,
as dictated by the EFT philosophy \cite{polch}. The previous first quantized theory
that we started with could then be disregarded, as well as the discussion of 
the steps that yielded the EFT (for further details on this procedure, please see,{\it e.g.}, equations (18) and
(19) in \cite{clz} or Section 2 in \cite{flsz}).

We now discuss the $\winf$ bosonization of the hamiltonian (\ref{series}). The 
quadratic form (upon use of the usual Sugawara formula for the Virasoro modes) in the RHS of (\ref{hcv1}) can
be diagonalized  by the following
Bogoliubov transformation
\barr
\w^0_{\ell}&=&\v^0_{\ell}\ \cosh \beta + \vb^0_{-\ell}\
\sinh \beta ~~~, \nl
\wb^0_{\ell}&=&\v^0_{-\ell}\ \sinh \beta +
\vb^0_{\ell}\ \cosh \beta
\label{bogo}
\earr
for all $\ell$, with $\tanh 2\beta =g/(2+g)$ for small $g$.
In fact, by replacing in (\ref{hcv1}) the inverse transformation to (\ref{bogo}),
expressing the $V^0_n$ operators in terms of the $W^0_n$ ones, we verify by explicit
computation that the (chirality) mixed terms with $W^0_n{\overline W}^0_n$ operators cancel out
if the condition $g\left[ \cosh^2(\beta )+  \sinh^2(\beta ) \right]
-2(2+g) \sinh(\beta )\cosh(\beta ) = 0 $ is met, which is equivalent to the condition
expressed above. With this value of $\beta$, the coefficient of the diagonal terms in
(\ref{hcv1}) becomes $\exp (2\beta) \simeq \xi $. 
The same procedure applied to (\ref{hcv0}) yields a term proportional to $W^0_0 + 
{\overline W}^0_0$ with a coefficient proportional to $\xi^{3/2}$. Finally, 
a similar treatment of (\ref{hcv2}) produces a number of terms. The cubic ones,
that we understand to be normal-ordered, yield mixed as well as diagonal terms. 
The mixed one has a coefficient proportional to 
$\left[ \cosh(\beta)-\sinh(\beta)\right] \left[ g/4 \left( \cosh^2(\beta )+  \sinh^2(\beta ) \right)
-(1+g/2) \sinh(\beta )\cosh(\beta ) \right]$, which vanishes automatically as the second factor is zero
due to condition that diagonalizes (\ref{hcv1}) \cite{cfslz,flsz}.
This result implies the {\it automatic} chiral factorization of the EFT of the CS model beyond
the $1/N$ accuracy, an unanticipated property that characterizes this theory \cite{flsz}. 
The coefficient of the diagonal term is 
$(1+g/4)\left[ \cosh^3(\beta )-  \sinh^3(\beta ) \right]
-(3g/4)\left[ \sinh(\beta) \cosh^2(\beta )-  \sinh^2(\beta ) \cosh(\beta) \right] = \exp(\beta) = \xi^{1/2} $.
Finally, the transformation parameters in (\ref{bogo}) are given by $\cosh(\beta)=(\xi + 1)/(2\sqrt{\xi})$ and
$\sinh(\beta)=(\xi - 1)/(2\sqrt{\xi})$.

The new basis of bosonic $\winf$ operators describes interacting fermions, and 
the corresponding ground state $|\ \Omega\ \rangle$ is a highest-weight state 
with respect to the redefined $\winf $ operators, namely
$\w^i_\ell |\ \Omega\ \rangle = 0$, $\ell > 0, i \geq 0$.
Notice that $|\ \Omega\ \rangle$ is a highly correlated state in terms
of the free fermionic vaccum $|\ \Omega\ \rangle_0$, a result implied
by the Bogoliubov transformation as well. 
The effective hamiltonian is the following operator: 
\barr
{\cal H}_{CS} &=&\left(
2\pi n_0  \sqrt{\xi}\right)^2
\left\{\left[\frac{\sqrt{\xi}}{4}\,\w_0^0
+\frac{1}{N}\,\w_0^1+
\frac{1}{N^2}\left(\frac{1}{\sqrt{\xi}}\,\w_0^2
-\frac{\sqrt{\xi}}{12}\,\w_0^0 \right.\right.
\right.\nl
&&-\ \left.\left.\left.
\frac{g}{2\xi^2}\,\sum_{\ell=1}^\infty
\,\ell~\w_{-\ell}^0\,\w_\ell^0\right)
\right]+\left(\,W~\leftrightarrow~{\overline W}\,\right)
\right\}~~~,
\label{hcsf}
\earr
where $\xi=\left(1+\sqrt{1+2g}\right)/2$ is the parameter defined after (\ref{ham}).
Note that $\xi = 1$ corresponds to the free fermion case.
The operators $~\w_{\ell}^m$
in (\ref{hcsf}) are the lowest (in $m=i+1$, where $i$ is the conformal spin \cite{bpz}) 
generators of the $\winf$ algebra. The terms in the $~\w_{\ell}^m$ ($\wb^i_\ell$) operators describe the 
dynamics at the right $(R)$ (left $(L)$) Fermi point, respectively. 
We remark that the {\it complete factorization} of (\ref{hcsf}) into chiral
and antichiral sectors is a consequence of the Bogoliubov
transformation (\ref{bogo}) that decouples both sectors, generically mixed 
by backward scattering terms in the fermionic form of the hamiltonian obtained 
in the thermodynamic limit \cite{cfslz,flsz}.
Ultimately, the disappearance of the terms that mix chirality in the low energy EFT is expected
at the $1/N$ level of accuracy,
as the Fermi momentum provides a large energy scale. For the same reason, a factorization 
is consistent with the absence of an energy scale and, therefore, conformal invariance.
The fulfillment of this result beyond the $1/N$ term is not automatic in general (see \cite{flsz} for
an extended discussion and further examples), but it 
is met in the theory under consideration.

We now turn our attention to the attractive four-fermion interaction of (\ref{hcv2}), 
considering the EFT as independent from the first quantized CS model and a {\it bona-fide} theory in its
own right, following \cite{polch}.
Conveniently normal ordered (dropping a constant term), it is given by:
\beq
{\cal H}^{att}_{(2)} 
=-\ \frac{g}{2}\left(\frac{2\pi}{N}\right)^2\,\sum_{\ell=0}^{\infty} \ell 
\left( \v^0_{-\ell}\,\v^0_{\ell}+
\vb^0_{-\ell} \,\vb^0_{\ell}+\v^0_{\ell} \,\vb^0_{\ell}\right) ~~~.
\label{hcv2bbno}
\eeq
The third term dominates in the UV region of large momenta because it involves
backward scattering processes, of the order of the Fermi momentum,  
It is expected to correspond to a short-range potential in coordinate 
space, given that, {\it e.g.}, the 2D Haldane psedupotentials in the QHE have 
four-fermion matrix elements that grow as $\ell^2 $ (for $\ell \gg 1$) \cite{wref}. 
To show it, we consider the interaction matrix elements ($k > 0$ is the continuous variable that stands for 
$|{\ell}|$) of the chiral sector
\beq
\lambda (k)\ =-\  \frac{g}{2}\left(\frac{2\pi}{N}\right)^2\ k
\label{lambdas}
\eeq
Inverse Fourier transforming these coefficients we expect to find  
an effective potential interaction in $x$ coordinate space ($x = R \theta$ is the coordinate along the 
circumference of
radius $R=L/(2\pi )$ and represents here the relative coordinate between two nearby particles). 
The interacting particles are bosons (particle-hole pairs) that move on opposite directions. 
\beq
\lambda (x)\ =\ \frac{1}{2\pi}\int_{-\infty}^{\infty}\ \lambda( k)\ e^{ikx}\ =\ 
-\  \frac{g}{2}\left(\frac{2\pi}{N^2}\right)\frac{\partial}{i\partial x }\int_{0}^{\infty}\ e^{ikx}\ .
\label{lambdax}
\eeq
We obtain
\beq
V_{eff}(x)\ =\ \frac{N^2}{\pi} \lambda (x)\ =\  
g\ \left[ \frac{1}{x^2}\ +\ i \pi \delta^{'}(x)\ \right]\ ,
\label{lambdax2}
\eeq
where prime denotes derivative with respect to the argument. 
The first term of (\ref{lambdax2}) coincides with the interaction in (\ref{ham})
for small $x$ and expresses the projection of the 3D $1/r^2$ potential
onto the 1D space of the circumference. 
The second term is a short range, 'dipolar' pseudopotential of the form 
\beq
V_d(x)\ =\ ig\  \pi \delta^{'}(x)\  ,
\label{vd}
\eeq
Pseudoptentials of this type have been studied (see {\it e.g.}, \cite{ttw}).
However, in this case it is imaginary, rendering the potential non-hermitian
and signaling time instability. 
Although not necessary at this point, one could trace back the origin of this term to the regularization 
conditions on the matrix elements of (\ref{hcv2bbno}) (see \cite{cfslz,flsz}),
such that all momenta are bounded by a UV cutoff that is removed when taking the
thermodynamic limit. As stated before, the EFT stands by itself and needs not make
reference to the steps that ended up with its formulation and this comment is meant
to illustrate the need for regularizations in EFTs \cite{polch}.
Starting from a 3D $1/r^2$ potential, 
the definition of the 1D EFT, in terms of the $x$ coordinate, 
demands further specification: it is required by the monodromy that 
defines quantum statistics and convenient for describing opposite moving 
particles. Quantum regularization 
in the spatial domain could be achieved by enlarging the circumference radially
onto a ribbon by point-splitting techniques.
In an infinitesimal arc of the circumference, the chord between its extremes
provides a physical realization of the ribbon thickening of the space
domain. In local complex coordinates, if $r$ denotes a small segment on the horizontal axis and $x$
the piece of arc on the upper half plane of almost the same length, 
the condition $r=x-i\varepsilon$ ($\varepsilon \to 0^{+}$ ) 
makes sense to specify a single-valued branch ( 
$\varepsilon = \Delta r$). Therefore, the existence of a pseudopotential of the form of (\ref{vd})
is implied by the workout of the Sokhotski-Plemelj relation \cite{grash}:
\beq
\frac{1}{x \pm i\varepsilon}= {\mathcal P} \left ( \frac{1}{x} \right )\ \mp i\pi \delta (x)\  ,
\eeq
where $\mathcal {P}$ denotes the Cauchy principal value and the limit 
$\varepsilon \to 0^{+}$ is understood.
Summing up, we have argued that the main physical effects of the CS interaction 
could be mainly attributed to the interaction (\ref{hcv2bbno})(\ref{lambdax2}).

Note that the $1/r^2$ potential yields an irrelevant 
interaction in 3D, as implied by the Landau theory \cite{polch}. However,
irrelevant interactions could give rise to interesting 
physical phenomena, like in the Fermi theory of beta decay.
We consider that in the CS model, interaction (\ref{vd}) between opposite particle-hole movers is a likely mechanism
that induces the transition to a highly correlated, bosonized 
EFT displaying charge and quantum statistics fractionalization. 

One could ask whether the CS quantum fluid describes particle behavior
in real samples. From (\ref{hcv2bbno}) we 
can hypothesize that the four fermion contact interaction
could be boson (such as a phonon) mediated.
The new interaction should also be planar, so that it would be expected
in strongly interacting systems with quasi 2D lattice backgrounds.
A transverse boson-particle momentum exchange, such as 
the angular momentum of a polarized boson, would be also necessary. This mechanism
and others deserve, in our opinion, further investigation. 

{\it In memory of Daniel R. Bes.} 
%
\def\NP{{\it Nucl. Phys.\ }}
\def\PRL{{\it Phys. Rev. Lett.\ }}
\def\PL{{\it Phys. Lett.\ }}
\def\PR{{\it Phys. Rev.\ }}
\def\IJMP{{\it Int. J. Mod. Phys.\ }}
\def\MPL{{\it Mod. Phys. Lett.\ }}

\end{document}